\documentclass{pasj00}
\draft

\usepackage{natbib}
\usepackage{threeparttable}

\begin{document}
\SetRunningHead{Yuan et. al.}{MIR data of AKARI/IRC as SFR indicators}

\title{AKARI/IRC Broadband Mid-infrared data as an indicator of Star Formation Rate}


%
 \author{%
   Fang-Ting \textsc{Yuan}\altaffilmark{1}, 
   Tsutomu~T. \textsc{Takeuchi}\altaffilmark{1}, 
   V\'{e}ronique \textsc{Buat}\altaffilmark{2}, 
   S\'{e}bastien \textsc{Heinis}\altaffilmark{2}, 
   Elodie \textsc{Giovannoli}\altaffilmark{2}, 
   Katsuhiro\ L. \textsc{Murata}\altaffilmark{1}, 
   Jorge \textsc{Iglesias-P\'{a}ramo}\altaffilmark{3,4}
   and
   Denis \textsc{Burgarella}\altaffilmark{2}}
 \altaffiltext{1}{Division of Particle and Astrophysical Sciences, Nagoya University, Furo-cho, Chikusa-ku, Nagoya 464--8602, JAPAN}
 \email{yuan.fangting, takeuchi.tsutomu, murata.katsuhiro@g.mbox.nagoya-u.ac.jp}
 \altaffiltext{2}{Laboratoire d'Astrophysique de Marseille, OAMP, Universit\'e Aix-Marseille, CNRS, 38 rue Fr\'ed\'eric Joliot-Curie, 13388 Marseille cedex 13, FRANCE}
 \email{veronique.buat, sebastien.heinis, elodie.giovannoli, denis.burgarella@oamp.fr}
 \altaffiltext{3}{Instituto de Astrof\'{\i}sica de Andaluc\'{\i}a (IAA - CSIC), Glorieta de la Astronom\'{\i}a s.n., 18008 Granada, SPAIN}
 \altaffiltext{4}{Centro Astron\'{o}mico Hispano Alem\'{a}n, C/ Jes\'{u}s Durb\'{a}n Rem\'{o}n 2-2, 04004 Almer\'{\i}a, SPAIN}
 \email{jiglesia@iaa.es}

\maketitle

\begin{abstract}
AKARI/Infrared Camera (IRC) Point Source Catalog provides a large amount of flux data at {\it S9W} ($9\ {\rm \mu m}$) and {\it L18W} ($18\ {\rm \mu m}$) bands. With the goal of constructing Star-Formation Rate(SFR) calculations using IRC data, we analyzed an IR selected GALEX-SDSS-2MASS-AKARI(IRC/Far-Infrared Surveyor) sample of 153 nearby galaxies. The far-infrared fluxes were obtained from AKARI diffuse maps to correct the underestimation for extended sources raised by the point-spread function photometry. SFRs of these galaxies were derived by the spectral energy distribution fitting program CIGALE. In spite of complicated features contained in these bands, both the {\it S9W} and {\it L18W} emission correlate with the SFR of galaxies. The SFR calibrations using {\it S9W} and {\it L18W} are presented for the first time. These calibrations agree well with previous works based on Spitzer data within the scatters, and should be applicable to dust-rich galaxies.   
\end{abstract}

\section{Introduction}
The star formation activity is fundamental to studies on the formation and evolution of galaxies. Numerous efforts have been made to find reliable and convenient SFR indicators \citep[e.g.][]{kenn98,hira03,bell03,hopk03,calze07}. Among most frequently used indicators, the ultraviolet (UV) and the optical recombination lines (e.g. ${\rm H}\alpha$, ${\rm Pa}\alpha$) give direct measures of light from young stars. However, UV and ${\rm H}\alpha$ emissions are strongly affected by dust extinction \citep{kenn98}, and ${\rm Pa}\alpha$ as well as other optical recombination lines were also shown to underestimate SFRs for galaxies with high luminosity \citep{rieke09}. On the other hand, since dust absorbs UV/optical light and re-emits the bulk of the energy into far-infrared (FIR) band ($25\ \mu{\rm m}\ \sim\ 350\mu{\rm m}$), FIR emission could efficiently trace SFRs for dusty galaxies. However, FIR emission is unable to trace the dust-unobscured radiation and includes part of radiation from old stellar populations. Therefore the energy balance method combining the FIR and UV derived SFRs were used to complement the emission from young stars not traced by FIR \citep[e.g.][]{bx96, meur99, gord00, buat05}. Although this method could trace SFRs with considerable accuracy, it is difficult to obtain the total dust emission especially for high redshift objects.  

The MIR monochromatic fluxes were also investigated as SFR indicators. The MIR emission is contributed by several components, including the polycyclic aromatic hydrocarbon (PAH) features (prominent at $6.2$, $7.7$, $8.6$, $11.3$, $12.7$ and $17\ {\rm \mu m}$), the continuum by the stochastic heating of very small grains, silicate absorption at $9.7$ and $18\ {\rm \mu m}$, molecular hydrogen lines and fine-structure lines \citep{lp84, lege89, dese90, dl07, smith07, trey10}. The MIR-SFR relation was intensively studied using Spitzer IRAC $8\ {\rm \mu m}$ and MIPS $24\ {\rm \mu m}$ photometry data and spectral data \citep[e.g.][]{for04, calze05, calze07, kenn09, rieke09, trey10}. Nevertheless, there is still a debate about the reliability of MIR indicators because of the complicated features it contains. Attempts to combine the optical and IR indicators show that this combination could trace the SFR effectively, and notably reduce the scatter \citep[e.g.][]{calze07, zhu08, kenn09}. However, factors of the combination are different in each work.

The recently released AKARI/IRC Point Source Catalogue Version $\beta$-1 (hereafter IRCPSC) provides positions and fluxes of all-sky survey at {\it S9W} ($9\ {\rm \mu m}$) and {\it L18W} ($18\ {\rm \mu m}$) bands \citep{ishi10}. For numerou MIR data, it would be useful if there is a benchmark of SFR measurement. Comparing with Spitzer $8$ and $24\ {\rm \mu m}$ bands, the AKARI IRC {\it S9W} and {\it L18W} bands cover wider wavelength ranges, including silicate absorption features at both bands and emission contributed by large PAH molecules at the {\it L18W} band. This work is dedicated to investigate whether and to what degree AKARI broadband MIR data could trace SFRs, and how much the inclusion of silicate absorption and longer wavelength PAH features would be.  

Start from a sample with multi-wavelength observation flux, we derive the SFR for each galaxy by the spectral energy distribution (SED) fitting. MIR data were correlated with the SFRs to built the SFR calibrations. Then the results were compared with those from the Spitzer observation. The present paper is organized as follows: Section 2 introduces the GALEX-SDSS-2MASS-AKARI sample. Section 3 gives a short introduction to methods for calculating SFRs. The MIR-SFR relations and some comparisons between AKARI and Spitzer are reported and discussed in Section 4. Conclusions are given in Sections 5. 

\section{Data}

\subsection{Construction of the multi-wavelength sample}
We cross-identified\footnote{The searching radius is set at $36''$ considering the original sample's position accuracy.} IRCPSC with multi-wavelength data constructed by \citet{take10} to obtain a sample including MIR photometric measurements. The original sample was based on a selection by IRAS-PSC${\it z}$ \citep{saun00} and AKARI/FIS Bright Source Catalog (hereafter FISBSC) data \citep{yama08, yama09}, which means all these galaxies have considerable fluxes at FIR. Then these galaxies were cross-matched with observations by GALEX, 2MASS and SDSS. Fluxes at the UV band were obtained by performing specific aperture photometry so as not to shred GALEX images. A detailed description of the original sample can be found in \citet{take10}. 

\begin{figure}
  \begin{center}
    \FigureFile(80mm,80mm){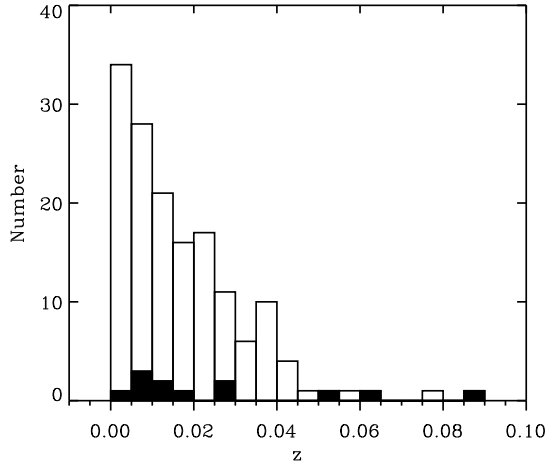}
  \end{center}
  \caption{The redshift distribution of our sample. The filled area denotes AGNs.}\label{fig:z_distri}
\end{figure}

IRCPSC presents flux data at $9\ \mu{\rm m}$ and $18\ \mu{\rm m}$ that have an effective bandwidth of $4.10\ \mu{\rm m}$ and a detection limit of $50\ {\rm mJy}$, and $9.97\ \mu{\rm m}$ and $90\ {\rm mJy}$, respectively \citep{ishi10}. After cross-identification, there were 162 galaxies with a flux of either $9\ \mu{\rm m}$ or $18\ \mu{\rm m}$. However, nine galaxies were found to have inconsistent fluxes at GALEX, SDSS or 2MASS band according to their SEDs, possibly due to the measurement errors or the misidentification of objects caused by the inhomogeneous resolution of each observation; they were omitted from our sample. We also discarded one galaxy with a too-small redshift ($z\ \sim\ 0.0008$). The summary of the sample is listed in Table \ref{tab:data_sum}. Searching in the SIMBAD database, most of galaxies in our sample are normal star forming galaxies: Only 15 galaxies are classified as AGNs (including Seyfert 1 and Seyfert 2 galaxies). All the galaxies in our sample are nearby ones. The distribution of the redshift is shown in Figure \ref{fig:z_distri}.

\subsection{The Re-estimation of AKARI FIR Data}

\begin{figure*}
  \begin{center}
    \FigureFile(150mm,150mm){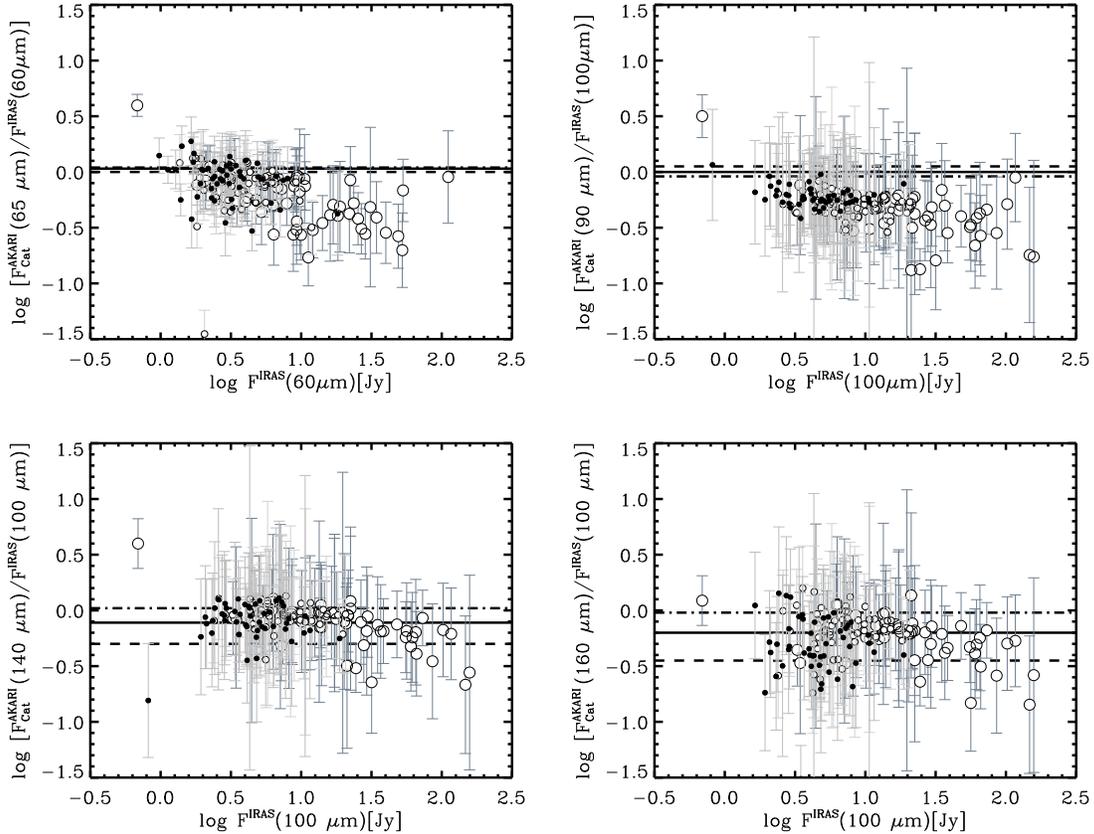}
  \end{center}
  \caption{Comparison between FIS catalog flux and the IRAS flux. Galaxies in different sub-sample are shown in different sizes (large open circles: large; small open circles: medium; dots: small). The DH models with three different $\alpha$ are shown in different lines (dashed:$\alpha=1.375$; solid: $\alpha=2.0$; dash-dotted: $\alpha=2.625$).}\label{fig:iras_cat}
\end{figure*}

The FISBSC flux density for extended sources would be no longer accurate because of the point source extraction procedure. In our sample, a considerable fraction of galaxies is the extended source. Therefore, it is questionable whether the catalog data of these sources are reliable. This is confirmed by making a comparison between FISBSC fluxes and IRAS co-added fluxes (Figure \ref{fig:iras_cat}), which were specially calculated for extended sources \citep{saun00}. In order to investigate the dependence of FISBSC flux on the extension of the galaxy, the sample was divided into three sub-samples according to the size of the galaxy (Table \ref{tab:div}). The lines in Figure \ref{fig:iras_cat} give predictions of the difference between two different bands by \citet{dh02} (hereafter DH) one-parameter constrained SED templates. The parameter $\alpha$ is related to IRAS flux ratio $f_{60}/f_{100}$. The $\log(f_{60}/f_{100})$ in our sample ranges from $-0.60$ to $0.25$, and most of the galaxies have $\log(f_{60}/f_{100})$ between $-0.5$ and $0.0$, corresponding to an $\alpha$ value between $2.625$ and $1.375$. Accordingly, the model prediction with $\alpha\ =\ 1.375$, $2.625$ and the median value $2.0$ are illustrated in Figure \ref{fig:iras_cat}. Table \ref{tab:data_comp} presents the median values of $\log(f_{\rm AKARI}/f_{\rm IRAS})$ for each sub-sample. Compared with the model predicted values, there are clear discrepancies between FISBSC and IRAS data, especially for $60-65\ \mu{\rm m}$ and $90-100\ \mu{\rm m}$. Also note that the large galaxies show greater discrepancies than small ones, indicating the PSF photometry is less reliable for larger galaxies, because a considerable part of their flux is left out by the relatively small beam size. 

Aiming to obtain more reliable flux data, the photometry of the diffuse maps provided by the AKARI group was conducted using SExtractor. The obtained ``AUTO'' fluxes were also compared with IRAS co-added fluxes. 

\begin{figure*}
  \begin{center}
    \FigureFile(150mm,150mm){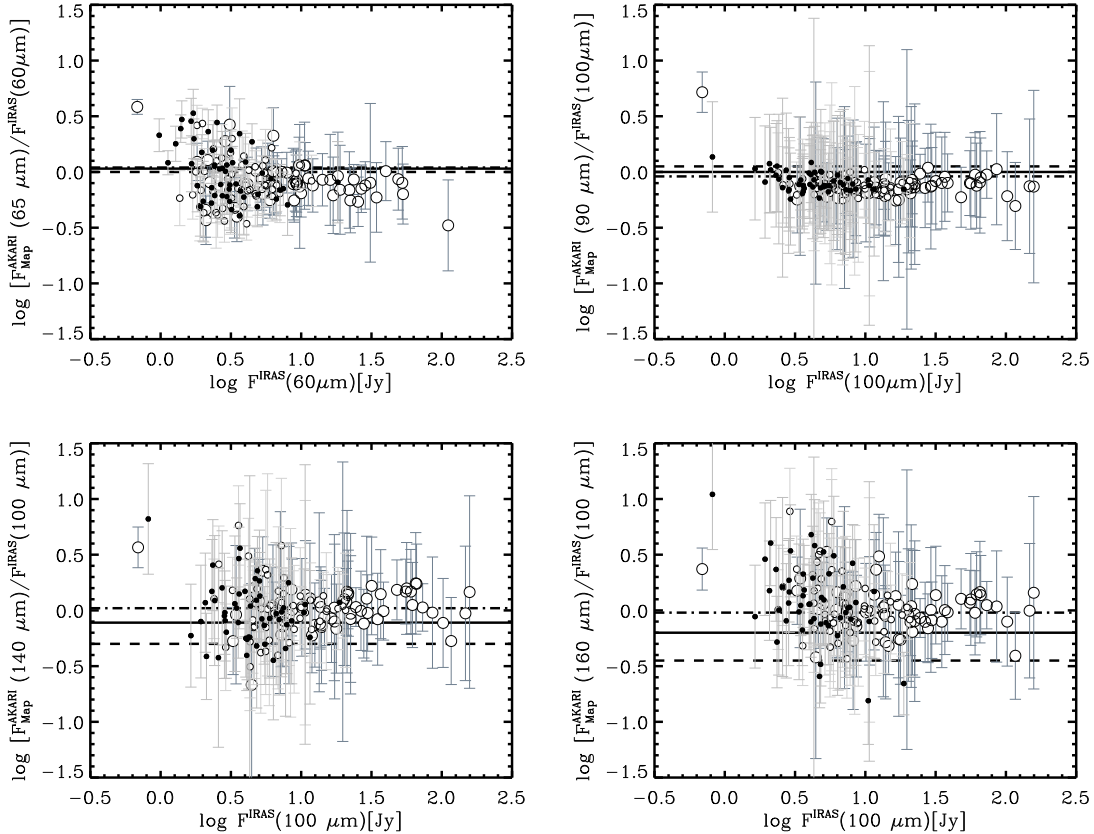}
  \end{center}
  \caption{Comparison between FIS diffuse map flux and the IRAS flux. Symbols are of the same meaning as in Figure \ref{fig:iras_cat}}\label{fig:iras_map}
\end{figure*}

In the $90\ \mu{\rm m}$ band, the result shows a great improvement on the consistency with the IRAS flux according to the DH model (Figure \ref{fig:iras_map} and Table \ref{tab:data_comp}) for all galaxies in our sample. At $65$ and $140\ \mu{\rm m}$ bands, the consistency is improved for those galaxies in `Medium' and `Large' subsamples, whereas the dispersion increases for the galaxies in 'Small' samples. Less improvement is found at $140\ \mu{\rm m}$ band, which can be explained by the relatively larger PSF FWHM at this band \citep[$\sim$ $60''$, as compared with $\sim$ $39''$ for the $65$ and $90\ \mu{\rm m}$ bands,][]{kawa07}. At the $160\ \mu{\rm m}$ band, neither fluxes derived from the diffuse maps nor those from the IRCPSC are satisfying, which is due to the poor quality of this band, thus $160\ \mu{\rm m}$ data were not used in the following work. Therefore, we kept IRCPSC flux values for `Small' sources at the $60$ and $140\ \mu{\rm m}$ bands, and applied fluxes derived from diffuse maps to the other sources. In addition, fluxes derived from AKARI diffuse maps showed notably smaller measurement errors than the IRAS fluxes, greatly improving the data quality. Considering taht  the IRAS fluxes provide information similar to that given by the AKARI bands, we omitted IRAS fluxes when fitting the SEDs.

Note that the MIR data provided by IRCPSC were little affected by an extension of the source, since it applies ``AUTO'' fluxes by SExtractor, which are suitable for both point sources and extended sources.

\section{SFR Calculation}

The SED fitting program CIGALE \citep{noll09} was used to calculate the SFR for our sample. CIGALE was developed to derive highly reliable galaxy properties by fitting the UV/optical SEDs and the related dust emission at the same time, i.e., the stellar population synthesized models are connected with infrared templates by the balance of the energy of dust emission and absorption. A detailed description of CIGALE can be found in \citet{noll09, buat10, giov10}. Here we give a brief introduction to the main features.

CIGALE allows one to use the stellar SEDs from models given either by \citet{mara05} or by \citet[][PEGASE]{fr97}. The difference between these models is the contribution of the thermally pulsating asymptotic giant branch (TP-AGB) stars. In PEGASE models, the contribution from TP-AGB stars is low \citep{mara06}. \citet{mara05} increased the contribution from TP-AGB stars adopting the 'fuel consumption' approach. \citet{mara06} shows that the insufficient consideration of TP-AGB stars overestimates the stellar mass by 0.2 dex and worsens the consistency with IR observation data at a redshift of $z\ \sim\ 2$. At lower redshift, it is found that using different models little affects the results \citep{rutt06, emin08}. Therefore, the models from \citet{mara05} were preferred in this work. The Kroupa IMF \citep{krou01} was used to calculate the complex stellar populations (CSPs). 

This code provides two scenarios of star formation: One is ``box models'' with a constant SFR, the other is ``$\tau$ models'', for which SFR decreases exponentially with a typical decay time, $\tau$. SFR is calculated as ${\rm SFR}_{\rm box}\ =\ M_{\rm gal}/t$ for ``box models'' and ${\rm SFR}_{\tau}\ =\ M_{\rm gal}/[\tau(e^{t/\tau}\ -\ 1)]$ for ``$\tau$ models'', where $M_{\rm gal}$ is the galaxy mass \citep{noll09}. CIGALE also allows one to apply different scenarios for young and old populations. The input SFH here is a constant burst SFH for young stellar populations, and an exponentially decreasing one for old stellar populations. Thus SFRs were calculated using the formula $f_{\rm ySP}\ \cdot\ {\rm SFR}_{box}\ +\ f_{\rm oSP}\ \cdot\ {\rm SFR}_{\tau}$, where $f_{\rm ySP}$ and $f_{\rm oSP}$ are fractions of young stellar populations and old stellar populations, respectively.       

The attenuation curve adopted by CIGALE is based on a law given by \citet{calze00}, with modification of the slope and/or adding a UV bump. The modification of the slope is controlled by the factor $(\lambda/\lambda_{V})^{\delta}$, i.e., by changing $\delta$, the slope of the attenuation curve can be modified. We only considered a modification of the slope of the attenuation law here, and no bump was introduced. CIGALE allows one to consider different effects of attenuation for old and young stellar populations by adding the reduction factor $f_{att}$ of the dust attenuation for the old stellar populations as an input parameter. For the IR part, CIGALE uses DH models, which is described in Section 2. Then the dust emission was calculated, and by balancing the energy emitted and absorbed, the short and long wavelength parts of the model were connected.

To compute the output parameters the code provides two methods: ``sum'' and ``max''. The former calculates the probability distribution functions (PDFs) by taking sums of the probability of the models in given bins of parameter space, which might cause an unintentional bias when the input parameter values are badly chosen. The latter introduces a fixed number of equally sized bins for each parameter and searches the maximum probability of models in each bin. Then these maximum probabilities are taken as weights for individual bins to calculate the expectation value of the parameter. The advantage of this method is that it alleviates the dependence on the choice of parameters \citep[see][for detail]{noll09}. Therefore, we applied the ``max'' method in this work.  

The input parameters applied in this study were adopted from \citet{buat10} in order to obtain a stable and reliable output. Since CIGALE is unable to trace the unobscured emission of an AGN, for Seyfert 1 galaxies, the output decreases the reliability \citep{buat10}. For dust-obscured AGNs, CIGALE provides models to fit the SED, and can avoid introducing any severe bias. Therefore, five Seyfert 1 galaxies were rejected in our sample, and other AGNs are marked during analysis. In order to keep the accuracy of derived SFRs, galaxies with relatively large discrepancies between observation and output spectra (reduced $\chi^2\ >\ 10$) were discarded as being unreliable. At first running, there were 25 such galaxies. By updating the SDSS data using Navigator of SDSS DR7/8 instead of the original pipeline data from SDSS DR7, the reduced $\chi^2$ values of 17 galaxies decreased to less than 10. \footnote{There is no modification for other sources which are fitted well.} Four of the eight remaining galaxies had very poor quality of AKARI diffuse maps. The other four had large extension and brightness, which could cause incomplete flux derivation or saturation in the optical plates. Considering that the number was small ($\sim$ 6\%), these eight galaxies and five out of the 17 galaxies were discarded. Therefore, reliable SFRs were derived for 140 galaxies, out of which there are 112 with available $9\ \rm{\mu m}$ fluxes and 97 with $18\ \rm{\mu m}$ fluxes.      

\subsection{The reliability of the results}

\begin{figure}
  \begin{center}
    \FigureFile(80mm,80mm){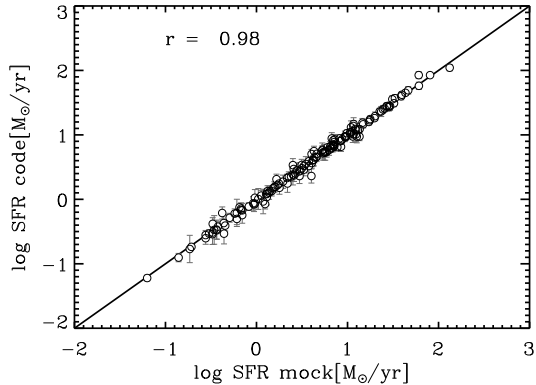}
  \end{center}
  \caption{The comparison between SFRs derived from CIGALE code and from the mock galaxies.}\label{fig:mock}
\end{figure}

For a SED fitting, the accuracy of the output depends on the input parameters. To give proper estimates of the SFRs, the robustness of the results must be tested. A straightforward way to check the reliability of the output of CIGALE is to use a sample of mock galaxies with comprehensively known physical parameters. Mock galaxies can be generated following a recipe in \citet{giov10}: 1, Run CIGALE on the data of real galaxies. For each galaxies, a best model is produced by $\chi^2$ minimization. Then, from these models, the fluxes at each bands can be estimated. 2, Add to each flux a random relative error, which is normally distributed with $\sigma\ =\ 0.1$. Thus, we obtain a mock catalog with flux information at every photometric band used in this study. The last step is to run the code on the mock catalog and then to compare the output parameters with the exact values provided by the best models. The result of the comparison is shown in Figure \ref{fig:mock}. We only present the result concerning SFRs because in this work the SFR is the only parameter that needs to be of concern. Figure \ref{fig:mock} shows the two quantities are well related, indicating SFRs derived here are reliable.  

\begin{figure}
  \begin{center}
    \FigureFile(80mm,80mm){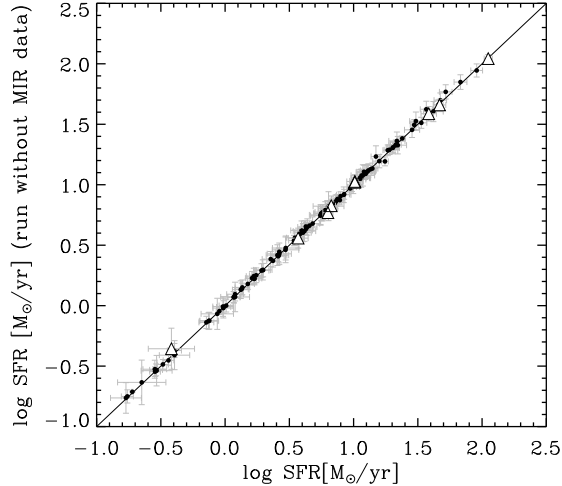}
  \end{center}
  \caption{The comparison between SFR derived with and without MIR data. AGNs are shown as triangles.}\label{fig:cmp_cig}
\end{figure}

As discussed in \citet{noll09}, CIGALE could provide stable results of SFRs as long as one constraint beyond PAH band is given. We ran CIGALE with and without MIR data to examine the influence on SFR by MIR photometric data. The result shows adding the MIR data or not adding has almost no influence on the resulting value of SFR (Figure \ref{fig:cmp_cig}), while not using FISBSC data brings large uncertainties to the output, consistent with the conclusion of \citet{noll09} using SINGS sample.

\section{Result and Discussion}

\begin{figure}
  \begin{center}
    \FigureFile(80mm,80mm){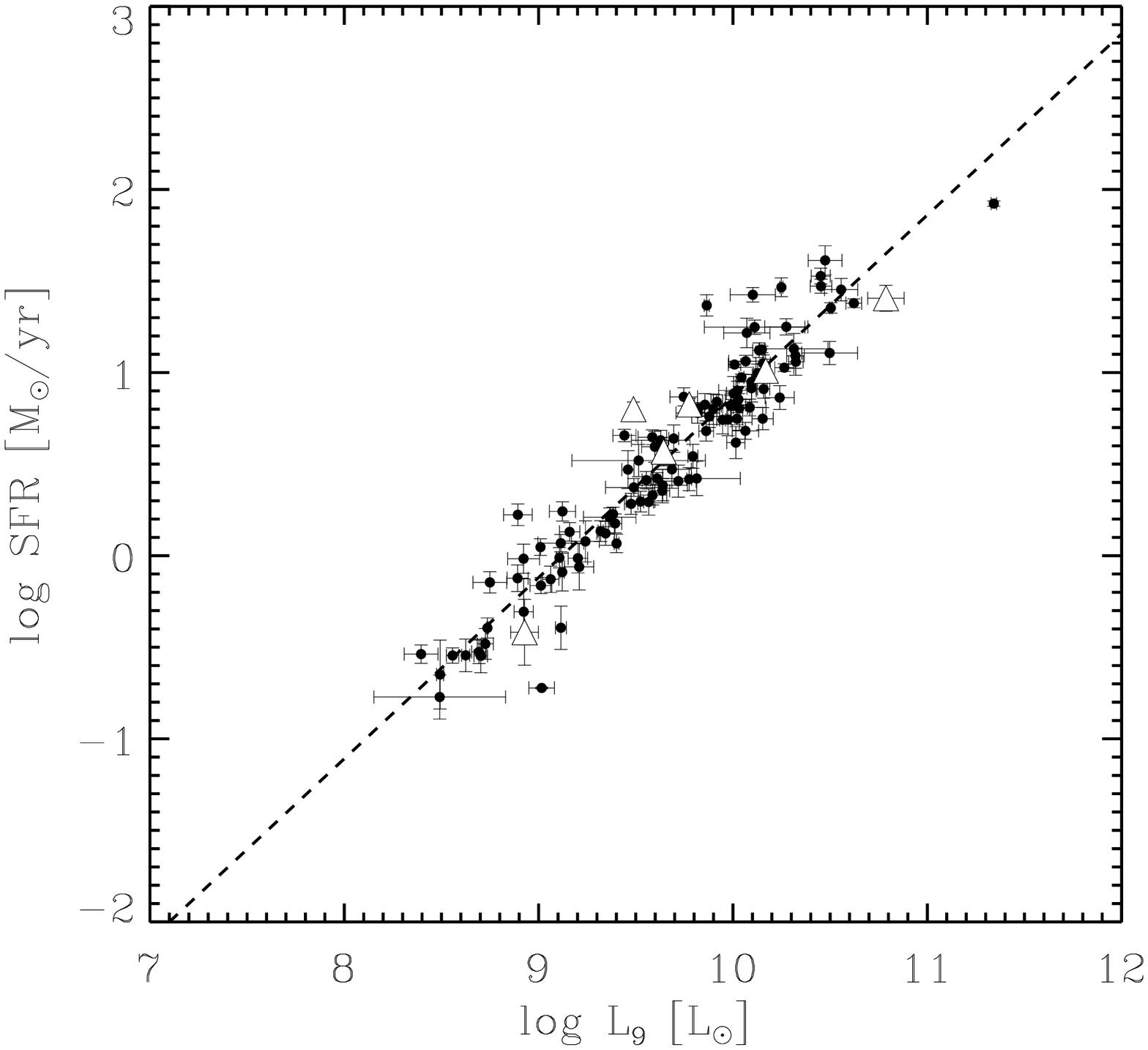}
  \end{center}
  \caption{The $9\ {\rm \mu m}$ luminosity-SFR relation. The dashed line shows the fitting result. The triangles are AGNs.}\label{fig:res09}
\end{figure}

\begin{figure}
  \begin{center}
    \FigureFile(80mm,80mm){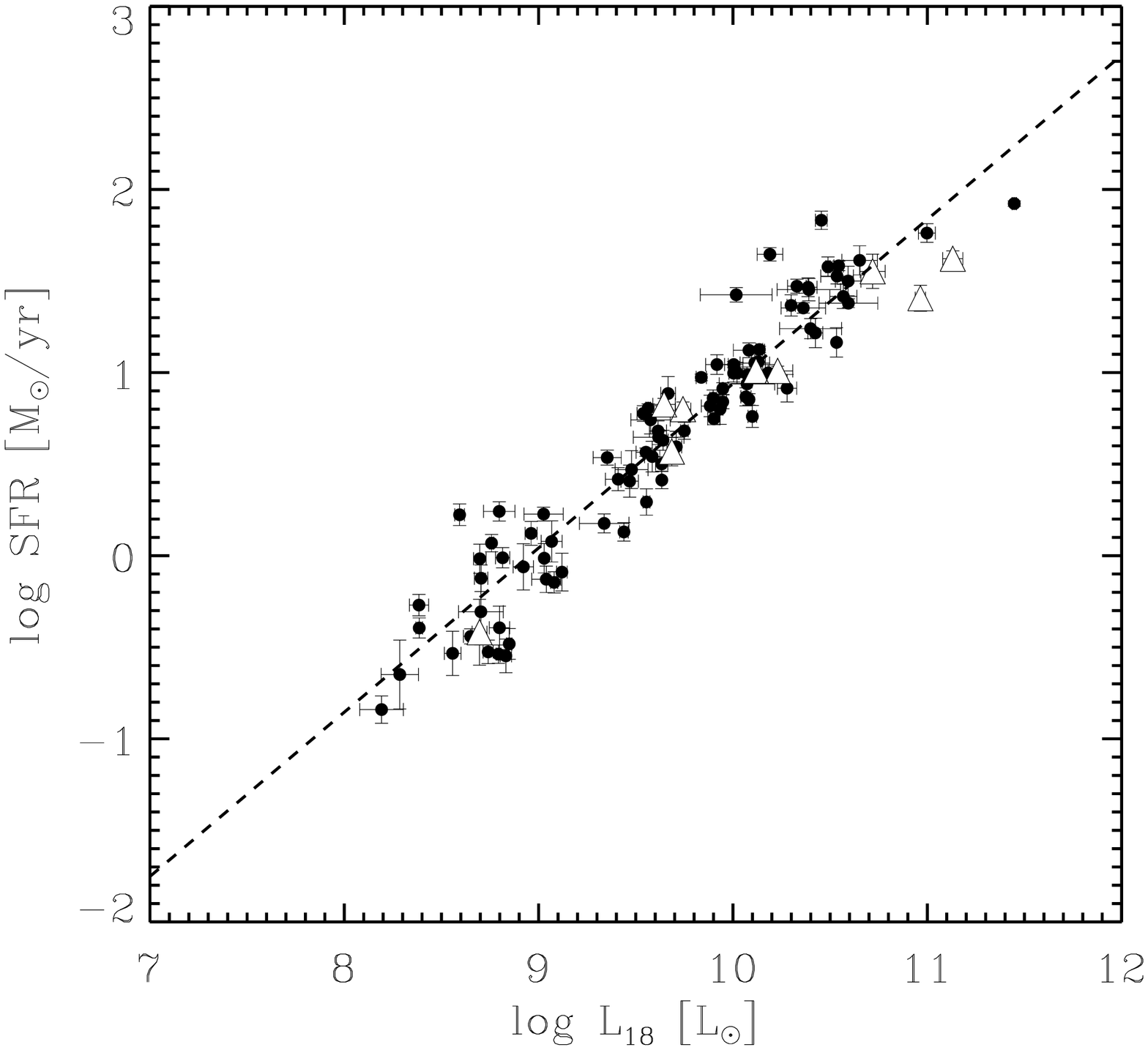}
  \end{center}
  \caption{The $18\ {\rm \mu m}$ luminosity-SFR relation. Lines and symbols share the same meaning with Figure \ref{fig:res09}.}\label{fig:res18}
\end{figure}

\begin{figure}
  \begin{center}
    \FigureFile(80mm,80mm){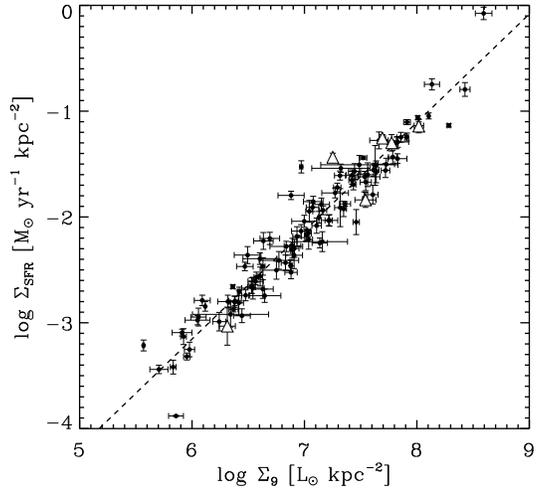}
  \end{center}
  \caption{The surface densities of $9\ {\rm \mu m}$ luminosity-SFR relation. The dashed line shows the fitting result. The triangles are AGNs.}\label{fig:res09_sf}
\end{figure}

\begin{figure}
  \begin{center}
    \FigureFile(80mm,80mm){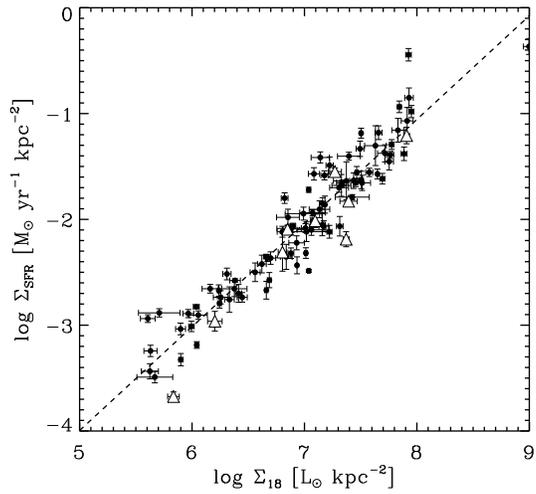}
  \end{center}
  \caption{The surface densities of $18\ {\rm \mu m}$ luminosity-SFR relation. Lines and symbols share the same meaning with Figure \ref{fig:res09}.}\label{fig:res18_sf}
\end{figure}

Apparently, the luminosity at $9\ \rm{\mu m}$ and $18\ \rm{\mu m}$ $L_{9}$ and $L_{18}$ \footnote{In this paper, $L_{\lambda}$ refers to $\nu L_{\nu}$ at wavelength $\lambda$.} both correlate with SFRs (Figure \ref{fig:res09} and Figure \ref{fig:res18}); the Spearman's correlation coefficients are 0.943 for the $L_{9}$-SFR relation and 0.956 for the $L_{18}$-SFR relation\footnote{Pearson's correlation coefficient is 0.944 for $L_{9}$-SFR relation and 0.951 for $L_{18}$-SFR relation}. Linear regressions following the method provided by \citet{kell07} give:
\begin{equation}
\log\ \frac{{\rm SFR}}{M_{\odot}/{\rm yr}}\ =\ (0.99\pm 0.03)\log\ \frac{L_{9}}{L_{\odot}} - (9.02 \pm 0.32)
\label{equ:res09}
\end{equation}
and  
\begin{equation}
\log\ \frac{{\rm SFR}}{M_{\odot}/{\rm yr}}\ =\ (0.90\pm 0.03)\log\ \frac{L_{18}}{L_{\odot}} - (8.03 \pm 0.30).
\label{equ:res18}
\end{equation}
The scatters of the data points about the regression lines of Equations \ref{equ:res09} and \ref{equ:res18} are approximately the same, with $\sigma\ =\ 0.18$ dex for $9\ \mu{\rm m}$ and $0.20$ dex for $18\ \mu{\rm m}$. These tight correlations also hold for the surface densities of luminosities and SFRs (Figures \ref{fig:res09_sf} and \ref{fig:res18_sf}). The correlation coefficients are 0.961 for $9\ \mu{\rm m}$ and 0.945 for $18\ \mu{\rm m}$. The areas of galaxies were calculated from {\it g}-band images of SDSS. The regression gives:
\begin{eqnarray}
\nonumber \log\ \frac{\Sigma_{\rm SFR}}{M_{\odot}{\rm yr}^{-1}{\rm kpc}^{-2}}\ &=&\ (1.02\pm 0.03)\log\ \frac{\Sigma_{9}}{L_{\odot}{\rm kpc}^{-2}} \\
&-&\ (9.30 \pm 0.19)
\label{equ:res09_sf}
\end{eqnarray}
with $\sigma\ =\ 0.18$, and  
\begin{eqnarray}
\log\ \frac{\Sigma_{\rm SFR}}{M_{\odot}{\rm yr}^{-1}{\rm kpc}^{-2}}\ &=&\ (0.98\pm 0.04)\log\ \frac{\Sigma_{18}}{L_{\odot}{\rm kpc}^{-2}} \\ 
&-&\ (8.89 \pm 0.25).
\label{equ:res18_sf}
\end{eqnarray}
with $0.22$.

Figures \ref{fig:res09} and \ref{fig:res18} show that AGNs share a MIR-SFR relation similar to normal star-forming galaxies, although the radiation mechanism of AGNs is different from normal galaxies. The slopes of the $\log L_{9}$-$\log$SFR and $\log L_{18}$-$\log$SFR relations derived here are almost equal to one, indicating the MIR-SFR relations are close to linear. 
\subsection{Comparison with SFR calibrations from Spitzer data}  
Spitzer $8\ \rm{\mu m}$ and $24\ \rm{\mu m}$ data were investigated as SFR tracers by several authors \citep[e.g][]{wu05, Perez06, calze07, relano07, zhu08, rieke09}. The $24\ \rm{\mu m}$ fluxes were found to be tightly related to the emission of the warm dust, and thus more intensively investigated, whereas the $8\ \rm{\mu m}$-SFR relation is more complicated, strongly depending on such as the metallicity, size and star-formation history \citep{calze07}, therefore, there are fewer calibrations.

\begin{figure}
  \begin{center}
    \FigureFile(80mm,80mm){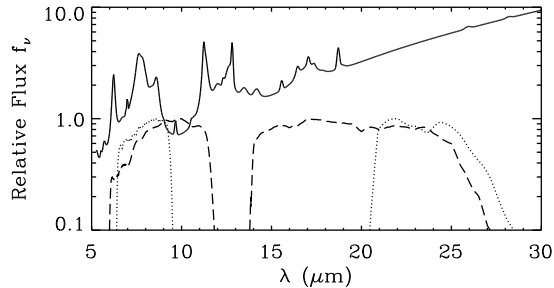}
  \end{center}
  \caption{The filter response curves of AKARI $9\ \rm{\mu m}$ and $18\ \rm{\mu m}$ bands (dashed line) and Spitzer $8\ \rm{\mu m}$ and $24\ \rm{\mu m}$ bands (dotted line). The solid line is the luminosity weighted average spectrum of star forming galaxies from \citet{smith07}.}\label{fig:filter_res}
\end{figure}

The main difference between the AKARI and Spitzer filters is their bandwidths. Due to the wider band, AKARI $9\ \rm{\mu m}$ and $18\ \rm{\mu m}$ fluxes are more affected by silicate absorption, PAH and molecular hydrogen line emissions (Figure \ref{fig:filter_res}). In order to check the reliability of our calibrations, the SFRs derived from Equations \ref{equ:res09} and \ref{equ:res18} were compared with those given by Spitzer calibrations. The work to compare was chosen to keep the luminosity range close to the present sample (Table \ref{tab:sfr_cal}). The calibrations given by \citet{wu05} and \citet{zhu08} are based on an equation given by \citet{kenn98} in which Salpeter IMF was used. The use of Salpeter IMF will cause smaller SFRs by $\sim\ 0.18$ dex than some other IMF with a more shallow slope at low masses \citep{rieke09}. Therefore this effect was corrected for the results of \citet{wu05} and \citet{zhu08}.

\subsubsection{Comparison between SFRs derived from $9\ \rm{\mu m}$ and from $8\ \rm{\mu m}$}
\begin{figure}
  \begin{center}
    \FigureFile(80mm,80mm){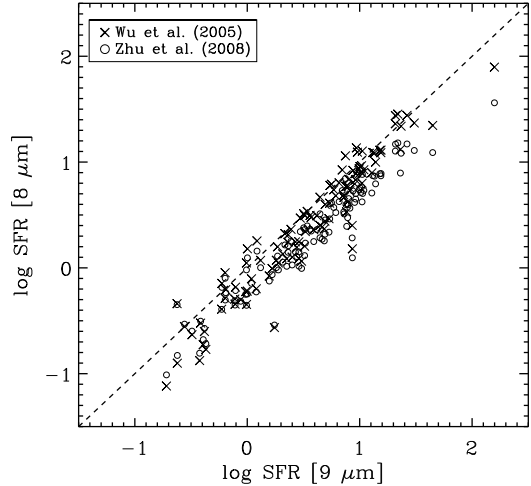}
  \end{center}
  \caption{Comparison between the SFRs derived from $9\ {\rm \mu m}$ emission (Equation \ref{equ:res09}) and from $8\ {\rm \mu m}$ emission by \citet{wu05} (crosses) and \citet{zhu08} (circles).}\label{fig:comp8_9}
\end{figure}

Spitzer $8\ \rm{\mu m}$ fluxes were computed from the output spectra of the CIGALE and filter response curves of Spitzer IRAC and MIPS. Since the $8\ \rm{\mu m}$ fluxes used in the calibration given in Table \ref{tab:sfr_cal} were the dust emission with the stellar contribution subtracted following the recipe of \citet{helo04}, the $3.6\ \mu{\rm m}$ flux was also calculated to compute the stellar composition contained in $8\ \rm{\mu m}$, and thus $8\ \rm{\mu m}$ dust emission could be obtained (hereafter, we refer to $8\ \rm{\mu m}$ dust emission as $8\ \rm{\mu m}$ emission for conciseness). Note that here the stellar contribution is very small, which can only affect the result by $\sim\ 0.02$ dex. The obtained flux is then converted to SFR by formula given by previous work (Table \ref{tab:sfr_cal}). The results are plotted in Figure \ref{fig:comp8_9} (The typical $1\sigma$ uncertainty for the galaxy of median luminosity is $\sim$ 0.5 dex). The statistical information of the comparison is given in Table 5.

The discrepancy between our results and \citet{wu05} may due to several reasons. \citet{wu05} listed factors such as the accuracy of fiber aperture corrections, the validity of the estimation of the obscuration in galaxies by using Balmer decrement, the possible contamination to radio and MIR emission from obscured weak AGNs. The larger capacity of our sample (79 for $8\ {\rm \mu m}$ in \citet{wu05} compared with 112 in our sample) and the wider coverage of $9\ {\rm \mu m}$ band may also cause such difference. Another possible reason is that the oversimplification of PAH emission in DH models underestimates the $8\ {\rm \mu m}$ flux and therefore gives smaller SFRs. However, this level of discrepancy is well within the scatters in Equations \ref{equ:res09}.

The discrepancy between \citet{wu05} and \citet{zhu08} is because \citet{zhu08} included $8\ {\rm \mu m}$-weak H${\rm II}$ galaxies with lower MIR luminosity \citep{zhu08}. Since no such galaxies were included in present sample, it is reasonable that present result agrees with \citet{wu05} better.

\subsubsection{Comparison between SFRs derived from $18\ \rm{\mu m}$ and $24\ \rm{\mu m}$}
\begin{figure}
  \begin{center}
    \FigureFile(80mm,80mm){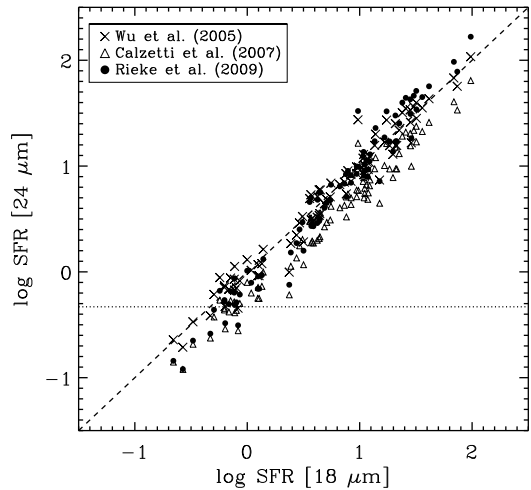}
  \end{center}
  \caption{Comparison between the SFRs derived from $18\ {\rm \mu m}$ (Equation \ref{equ:res18}) and from $24\ {\rm \mu m}$ emission by \citet{wu05} (crosses), \citet{calze07} (triangles) and \citet{rieke09} (dots). The dotted line gives the lower limit where the calibration of \citet{rieke09} applies.}\label{fig:comp18_24}
\end{figure}

For a $18\ {\rm \mu m}${--}$24\ {\rm \mu m}$ SFRs comparison, more reference calibrations are available (Table \ref{tab:sfr_cal}). The $24\ {\rm \mu m}$ fluxes are also derived from the output spectra of the SED fitting by CIGALE. The statistical information of the comparison is given in Table \ref{tab:18_24}. The converted SFRs are plotted in Figure \ref{fig:comp18_24}. Since the SFRs derived from \citet{wu05} are quite similar to those from \citet{zhu08} and the luminosity range in \citet{wu05} is closer to this work, the results from \citet{zhu08} are omitted for the sake of brevity. 

The results of \citet{wu05} and \citet{zhu08} agree well with our result after the correction of IMF. \citet{rieke09} assembled SED templates for local luminous and ultraluminous infrared galaxies and combined the result of \citet{dale07} and \citet{smith07} to produce templates at lower luminosities. Their result is applicable to galaxies with $24\ {\rm \mu m}$ luminosity greater than $6\ \times\ 10^8\ L_{\odot}$, corresponding to $\log\ {\rm SFR}\ =\ -0.33$, which is shown by the dotted line in Figure \ref{fig:comp18_24}. The present result agrees very well with \citet{rieke09} above the limit. 

Our result is a little higher than the one given by \citet{calze07}. A possible reason is that the result of \citet{calze07} was derived for H${\rm II}$ clouds by Pa$\alpha$ emission, which might be poorly applied to galaxy-wide calculations, because the diffuse MIR or Pa$\alpha$ emissions in the whole galaxy are not included \citep{alon06, calze07, kenn07, rieke09}.

\subsection{Combination of FUV and MIR indicators}
At lower IR luminosity, the IR indicator may fail to trace part of the UV photons from young stars due to the increased transparency of the ISM. A combination of unobscured FUV and MIR luminosities, (${\rm FUV}\ +\ \alpha{\rm MIR}$), may efficiently compensate for the lost energy, and could trace the SFR linearly \citep{zhu08}. However, upon converting SFRs to dust obscuration corrected FUV fluxes by eq.[1] from \citet{kenn98}, we find that $\alpha\ =\ 2.53$ and $3.33$ with a scatter of $0.17$ dex for $9\ \rm{\mu m}$ and $0.20$ dex for $18\ \rm{\mu m}$, respectively. The scatter is not reduced significantly. This fact indicates that the origin of the scatter in MIR-SFR diagram is complicated: not only the untraced UV photons, but also other unknown factors, such as the variation of the physical conditions within each galaxy, the distribution of dust and photo dissociation regions (PDRs), etc.    

\subsection{Metallicity}
We attempt to investigate the gas-phase metallicity range of our sample by searching in the metallicity database measured by \citet{trem04} for SDSS galaxies. Unfortunately, the $12\ +\ \log\rm{(O/H)}$ values are given for only 33 galaxies (all higher than 8.75). Therefore, we applied a compromised method: to investigate the stellar mass $M_{*}$ of our sample. Measured by CIGALE, the stellar mass $M_{*}$ of all the galaxy in the sample is larger than $10^{8.5}$ $M_{\odot}$. Thus, from mass-metallicity relation given by \citet{trem04}:
\begin{eqnarray}
  \nonumber 12\ +\ \log({\rm O/H})\ =\ &-&\ 1.492\ +\ 1.847(\log M_{*}) \\
  &-&\ 0.08026(\log M_{*})^2,
\end{eqnarray} 
the metallicity $12\ +\ \log\rm{(O/H)}$ for our sample is higher than $8.4$. This is not surprising because the initial sample from \citet{take10} is IR selected, which means a considerably high luminosity in IR and thus sufficient dust content and relatively high metallicity range. Thus, the MIR-SFR relations derived here could only be extrapolated to other high-metallicity galaxies. The situation for low metallicity galaxies is rather complicated; since the opacity of the galaxy decreases, MIR would be unable to trace most of the UV/optical photons and thus lose the ability as an SFR indicator \citep{calze07}. 

\subsection{AGNs}
\begin{figure}
  \begin{center}
    \FigureFile(80mm,80mm){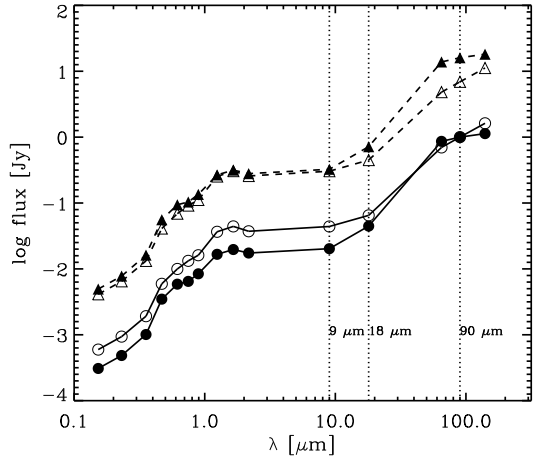}
  \end{center}
  \caption{The average SED (dashed lines) and the SED normalized at $90\ {\rm \mu m}$ (solid lines) of AGNs (solid symbols) and normal galaxies (open symbols).}\label{fig:ave_sed}
\end{figure}

Although AGNs have distinct features from normal galaxies in various physical properties, they share the same trend in the MIR-SFR figures. A possible reason is that the contribution from AGN component is minor (from SED fitting, less than 15\%), therefore the host galaxy component dominates the spectrum. To investigate the effect of AGNs on MIR emission, we plot average SEDs for AGNs and normal galaxies in Figure \ref{fig:ave_sed}. On average, AGNs are brighter than normal galaxies at all bands, while with a lower MIR/$90\ {\rm \mu m}$ ratio (Table \ref{tab:mir_90}). There are two possible reasons: 1, silicate absorption occurs more strongly in AGNs. 2, PAH molecules are destroyed by the harsh radiation field in AGNs. The latter is more convincing. Studies show that small PAH molecules contributed to shorter wavelength MIR emission are destroyed more easily than large ones contributed to longer wavelength MIR emission \citep[][and reference therein]{smith07, trey10}, which is consistent with the lower $9\ {\rm \mu m}$ flux value than $18\ {\rm \mu m}$ in Figure \ref{fig:ave_sed}. Nevertheless, these differences between AGNs and normal galaxies are only on their average level, which could not be used to distinguish AGNs and normal galaxies in individual cases. 

\section{Conclusion}

We combined AKARI/IRC $9\ {\rm \mu m}$ and $18\ {\rm \mu m}$ data with a previous sample to construct a FIR selected multi-wavelength sample with MIR photometric measurements. The FIR data of AKARI/FIS in the original sample were re-estimated by photometry of AKARI diffuse maps to correct the bias of PSF photometry for extended sources. Then, the SEDs of the sample were fitted by CIGALE, and the SFRs were obtained. Regression analysis was conducted to investigate MIR-SFR relations. SFRs converted from AKARI MIR fluxes were compared with those from the Spitzer MIR fluxes to test the reliability of AKARI MIR-SFR calibrations. From the result, we draw the following conclusions: 
\begin{enumerate}
\item Both $9\ {\rm \mu m}$ and $18\ {\rm \mu m}$ luminosities correlate with SFRs, and thus could be converted SFRs. 
\item A combination of FUV and MIR luminosity barely reduces the scatters, indicating that the unobscured UV photons are not the only reason of the variation of MIR-SFR relation. 

\item A comparison of the SFRs derived from Equations \ref{equ:res09} and \ref{equ:res18} with the ones derived from Spitzer MIR-SFR relations shows that the silicate absorption included in {\it S9W} ($9\ {\rm \mu m}$) and {\it L18W} ($18\ {\rm \mu m}$) bands little affects the results. The discrepancies, if any, are well within the uncertainties.

\item AGNs in the sample show no discrepancy with normal galaxies in the MIR-SFR diagrams. The smaller average MIR fluxes for AGNs than normal galaxies might indicate the small PAH molecules are destructed by harsh radiation from AGNs.
\end{enumerate}
In summary, for IR selected galaxies the rest frame $9\ {\rm \mu m}$ and $18\ {\rm \mu m}$ emissions are efficient tracers of SFRs, and the equations derived here should be applicable to other dust rich galaxies.

\bigskip

This work is based on observations with AKARI, a JAXA project with the participation of ESA. FTY, TTT, and KLM are partially supported from the Grand-in-Aid for the Global COE Program ``Quest for Fundamental Principles in the Universe: from Particles to the Solar System and the Cosmos'' from the Ministry of Education, Culture, Sports, Science and Technology (MEXT). TTT has been supported by Program for Improvement of Research Environment for Young Researchers from Special Coordination Funds for Promoting Science and Technology, and the Grant-in-Aid for the Scientific Research Fund (20740105) commissioned by the MEXT of Japan. VB and DB have been supported by the Centre National des Etudes Spatiales (CNES) and the Programme National Galaxies (PNG). JIP is supported by the grant AYA2007-67965-C03-02, from the Spanish MICINN. This work makes use of the SIMBAD database, operated at the Centre de Donées astronomiques de Strasbourg (CDS), Strasbourg, France. We acknowledge Y. Doi for preparing diffuse maps for our sample and Y. Matsuoka for helpful advice and comments. We also thank the anonymous referee for constructive suggestions to improve this paper. 



\newpage
\newpage
\begin{table*}
  \caption{A brief summary of the sample. Data with an asterisk (*) are not used for SED fitting in Section 3.}\label{tab:data_sum}
  \begin{center}
    \begin{tabular}{cccc}
      \hline
      Survey & Band & Wavelength ($\mu{\rm m}$)& Nb. of sources\\
      \hline
      GALEX & FUV, NUV & $0.153$, $0.231$ & 153\\
      SDSS  & {\it u}, {\it g}, {\it r}, {\it i}, {\it z} & $0.355$, $0.469$, $0.617$, $0.748$, $0.893$ & 153\\
	  2MASS & {\it J}, {\it H}, {\it Ks} & $1.244$, $1.655$, $2.169$ & 153   \\ 
	  IRAS* & band-1, 2, 3, 4 & $12$, $25$, $60$, $100$ & 153 \\     
      AKARI IRC & {\it S9W} & $9$ &  126 \\
      AKARI IRC & {\it L18W} & $18$ &  106\\
      AKARI FIS & {\it N60}, {\it Wide-S}, {\it Wide-L}, {\it N160}* & $65$, $90$, $140$, $160$ & 153\\ 
      \hline
    \end{tabular}
  \end{center}
\end{table*}

\begin{table*}
  \caption{Criteria to divide the sample into three sub-samples. The length of major axis $a$ and minor axis $b$ of each galaxy is obtained from the SDSS image. The $40''$ is taken as the separating value considering the PSF size of AKARI/FIS.}\label{tab:div}
  \begin{center}
    \begin{tabular}{cc}
      \hline
      Sub-sample & Criteria\\
      \hline
      Small   & $a\ <\ 40''$ and $b\ <\ 40''$\\
      Medium  & $a\ >\ 40''$ or $b\ >\ 40''$\\
	  Large   & $a\ >\ 40''$ and $b\ >\ 40''$\\      
      \hline
    \end{tabular}
  \end{center}
\end{table*}

\begin{table*}
  \caption{Comparison between AKARI and IRAS flux. The mean values of the difference between AKARI and IRAS fluxes according to AKARI FISBSC fluxes (Catalog), fluxes derived from diffuse maps (Map) and DH model prediction (Model) are listed for each subsample of galaxies ('L', 'M' and 'S' represent large, medium and small group in Table \ref{tab:div}, respectively).}\label{tab:data_comp}
  \begin{center}
    \begin{tabular}{c ccc ccc ccc}
      \hline
        & \multicolumn{3}{c}{Catalog} & \multicolumn{3}{c}{Map} & \multicolumn{3}{c}{Model}\\[0.5ex]
      \hline
         & L & M & S & L & M & S & $\alpha\ =\ 1.375$ & $\alpha\ =\ 2$ & $\alpha\ =\ 2.625$ \\[0.5ex]
      \hline
	  $\log(f_{65}/f_{60})$ & -0.18 & -0.09 & -0.03 & -0.10 & -0.12 & -0.01 & 0.00 & 0.03 & 0.04\\
	  $\log(f_{90}/f_{100})$ & -0.33 & -0.27 & -0.24 & -0.13 & -0.12 & -0.11 & 0.05 & 0.00 & -0.04\\
	  $\log(f_{140}/f_{100})$ & -0.09 & -0.04 & -0.09 & 0.03 & 0.01 & -0.02 & -0.30 & -0.11 & 0.02\\
	  $\log(f_{160}/f_{100})$ & -0.20 & -0.17 & -0.32 & 0.01 & 0.00 & 0.07 & -0.45 & -0.20 & -0.02\\[0.5ex]
      \hline
    \end{tabular}
  \end{center}
\end{table*}

\begin{table*}
  \caption{SFR calculations based on Spitzer data. The luminosity is expressed in $L_{\odot}$ and SFR in $M_{\odot}/{\rm yr}$.}\label{tab:sfr_cal}
\centering
\begin{threeparttable}
    \begin{tabular}{ll}
      \hline
      Work & SFR calculation\\[0.5ex]
      \hline
      \citet{wu05}  & $\log {\rm SFR}\ =\ (1.09\ \pm\ 0.06)\log L_{8}\ -\ (10.03\ \pm\ 0.16)$\\[1ex]
      \citet{zhu08}  & $\log {\rm SFR}\ =\ (0.93\ \pm\ 0.03)\log L_{8}\ -\ (8.59\ \pm\ 0.08)$\\[0.5ex]
	  \hline      
      \citet{wu05}  & $\log {\rm SFR}\ =\ (0.89\ \pm\ 0.06)\log L_{24}\ -\ (7.82\ \pm\ 0.17)$\\[1ex]
      \citet{calze07}\tnote{*}  & $\log {\rm SFR}\ =\ 0.8850\log L_{24}\ -\ 8.17$ (1 $\sigma$ uncertainty 0.03)\\[1ex]
      \citet{zhu08}  & $\log {\rm SFR}\ =\ (0.85\ \pm\ 0.01)\log L_{24}\ -\ (7.47\ \pm\ 0.06)$\\[1ex]
       & ${\rm SFR}\ =\ 7.8\times10^{-10}L_{24}$ for $6\times10^8 \le L_{24} \le 1.3\times10^{10}\ L_{\odot}$\\[-1ex]
      \raisebox{1.5ex}{\citet{rieke09}} & ${\rm SFR}\ =\ 7.8\times10^{-10}L_{24}\times(7.76\times10^{-11}L_{24})^{0.048}$ for $L_{24} > 1.3\times10^{10}\ L_{\odot}$\\[1ex]
      \hline
    \end{tabular}   
	\begin{tablenotes}\footnotesize
	\item[*] For galaxies with $12\ +\ \log({\rm O/H})\ >\ 8.35$, i.e., ``high metallicity'' galaxies in \citet{calze07}.
	\end{tablenotes}
\end{threeparttable} 
\end{table*}

\begin{table*}
  \caption{Regressions between SFRs derived from AKARI $9\ {\rm \mu m}$ and from Spitzer $8\ {\rm \mu m}$: $\log {\rm SFR}(8\ {\rm \mu m})\ =\ a\ +\ b\log{\rm SFR}(9\ {\rm \mu m})$ and the mean value of the difference: $<\log {\rm SFR}(8\ {\rm \mu m})-\log{\rm SFR}(9\ {\rm \mu m})>$.}\label{tab:8_9}
  \begin{center}
    \begin{tabular}{ccccc}
      \hline
       $8\ {\rm \mu m}$ calibration & $a$ & $b$ & Correlation coefficient & $<\log {\rm SFR}(8\ {\rm \mu m})-\log{\rm SFR}(9\ {\rm \mu m})>$\\[0.5ex]
      \hline
	  \citet{wu05} & $-0.17\ \pm\ 0.02$ & $1.05\ \pm\ 0.03$ & 0.968 & $-0.14\ \pm\ 0.18$\\
      \citet{zhu08} & $-0.20\ \pm\ 0.02$ & $0.90\ \pm\ 0.02$ & 0.968 & $-0.25\ \pm\ 0.16$\\[0.5ex]
      \hline
    \end{tabular}
  \end{center}
\end{table*}

\begin{table*}
  \caption{Regressions between SFRs derived from AKARI $18\ {\rm \mu m}$ and from Spitzer $24\ {\rm \mu m}$: $\log {\rm SFR}(24\ {\rm \mu m})\ =\ a\ +\ b\log{\rm SFR}(18\ {\rm \mu m})$ and the mean value of the difference: $<\log {\rm SFR}(24\ {\rm \mu m})-\log{\rm SFR}(18\ {\rm \mu m})>$.}\label{tab:18_24}
  \begin{center}
    \begin{tabular}{ccccc}
      \hline
       $24\ {\rm \mu m}$ calibration & $a$ & $b$ & Correlation coefficient & $<\log {\rm SFR}(24\ {\rm \mu m})-\log{\rm SFR}(18\ {\rm \mu m})>$\\[0.5ex]
      \hline
	  \citet{wu05} & $-0.02\ \pm\ 0.02$ & $1.00\ \pm\ 0.02$ & 0.982 & $-0.01\ \pm\ 0.12$\\
	  \citet{calze07} & $-0.23\ \pm\ 0.02$ & $1.00\ \pm\ 0.02$ & 0.982 & $-0.23\ \pm\ 0.12$\\
      \citet{zhu08} & $-0.02\ \pm\ 0.02$ & $0.96\ \pm\ 0.02$ & 0.982 & $-0.05\ \pm\ 0.11$\\
      \citet{rieke09} & $-0.14\ \pm\ 0.02$ & $1.14\ \pm\ 0.02$ & 0.982 & $-0.04\ \pm\ 0.17$\\[0.5ex]
      \hline
    \end{tabular}
  \end{center}  
\end{table*}

\begin{table*}
  \caption{The MIR to $90\ {\rm \mu m}$ flux ratio for AGNs and normal galaxies.} \label{tab:mir_90}
  \begin{center}
    \begin{tabular}{cccc}
      \hline
      Gal. & $\log (f_{9}/f_{90})$ & $\log (f_{18}/f_{90})$ & $\log (f_{9}/f_{18})$\\[0.5ex]
      \hline
      AGN   & -1.69 & -1.35 & -0.34\\[0.5ex]
      Normal  & -1.35 & -1.19 & -0.16\\[0.5ex]    
      \hline
    \end{tabular}
  \end{center}
\end{table*}
\end{document}